# Exoplanet Science Priorities from the Perspective of Internal and Surface Processes for Silicate and Ice Dominated Worlds

A white paper submitted in response to the National Academy of Sciences 2018 Exoplanet Science Strategy solicitation, from the NASA Nexus for Exoplanetary System Science (NExSS)

Wade G. Henning[1, 2 *], Joseph P. Renaud[3], Avi M. Mandell[1], Prabal Saxena[1], Terry A. Hurford[1], Soko Matsumura[4], Lori S. Glaze[1], Timothy A. Livengood[1,2], Vladimir Airapetian[1,5], Erik Asphaug[6], Johanna K. Teske[7], Edward Schwieterman[8], Michael Efroimsky[9], Valeri V. Makarov[9], Ciprian T. Berghea[9], Jacob Bleacher[1], Andrew Rushby[10], Yuni Lee[1], Weijia Kuang[1], Rory Barnes[11], Chuanfei Dong[12], Peter Driscoll[13], Shawn D. Domagal-Goldman[1], Nicholas C. Schmerr[2], Anthony D. Del Genio[14], Adam G. Jensen[15], Lisa Kaltenegger[16], Linda Elkins-Tanton[17], Everett L. Shock[17], Linda E. Sohl[18], Elisa Quintana[1], Laura Schaefer[17], Thomas S. Barclay[1,2], Yuka Fujii[19], Keiko Hamano[19], Noah E. Petro[1], Eric D. Lopez[1], Dimitar D. Sasselov[20]

**Author Affiliations:**
1. NASA Goddard Space Flight Center, 2. University of Maryland, 3. George Mason University, 4. University of Dundee, 5. American University, 6. Lunar and Planetary Laboratory, University of Arizona, Tucson, 7. Carnegie DTM, Washington DC, 8. University of California, Riverside, 9. United States Naval Observatory, 10. NASA Ames Research Center, 11. University of Washington, 12. Princeton University, 13. Carnegie Institution for Science, 14. NASA Goddard Institute for Space Studies, 15. University of Nebraska at Kearney, 16. Carl Sagan Institute Cornell University, 17. Arizona State University, 18. Columbia University, 19. Earth-Life Science Institute, Tokyo Institute of Technology, 20. Harvard University

* Corresponding Author: 301-614-5649, wade.g.henning@nasa.gov

**Acronyms:**

| | |
|---|---|
| NExSS | Nexus for Exoplanetary System Science |
| HST | Hubble Space Telescope |
| JWST | James Webb Space Telescope |
| WFIRST | Wide Field Infrared Survey Telescope |
| EPOXI | Extrasolar Planet Observation Deep Impact Extended Investigation |
| OST | Origins Space Telescope |
| HabEx | Habitable Exoplanet Imaging Mission |
| LUVOIR | Large Ultraviolet Optical and Infrared Surveyor |
| TESS | Transiting Exoplanet Survey Satellite |
| MMR | Mean Motion Resonance |
| SNR | Signal to Noise Ratio |
| RV | Radial Velocity |
| TTV/TDV | Transit Timing Variations/Transit Depth Variations |
| GCM | General Circulation Model |

**Introduction:**

The geophysics of extrasolar planets is a scientific topic that is often regarded as standing largely beyond the reach of near-term observations. This reality in no way diminishes the central role of geophysical phenomena in shaping planetary outcomes, from formation, to thermal and chemical evolution, to numerous issues of surface and near-surface habitability. We emphasize that for a balanced understanding of extrasolar planets, it is important to look beyond the natural biases of current observing tools, and actively seek unique pathways to understand exoplanet interiors as best as possible during the long interim prior to a time when internal components are more directly accessible. Such pathways include but are not limited to: a.) enhanced theoretical and numerical modeling development, b.) laboratory research on critical material properties, c.) measurement of geophysical properties by indirect inference from imprints left on atmospheric and orbital properties, and d.) the purpose-driven use of Solar System object exploration expressly for its value in comparative planetology toward exoplanet-analogs. Breaking down barriers that envision local Solar System exploration, as a wholly separate discipline, in financial competition with extrasolar planet research, may greatly improve the rate of needed scientific progress for exoplanet geophysics.

This whitepaper is submitted as part of a suite of works by NASA's Nexus for Exoplanetary System Science (NExSS) in response to the National Academy of Sciences Exoplanet Science Strategy request. Other NExSS-led whitepaper topics submitted include: biosignature detection (Domagal-Goldman et al.), exoplanet climates (Del Genio et al.), star-planet interactions (Airapetian et al.), planet formation (Apai et al.), stellar noise (Apai et al.), and exoplanet diversity (Kopparapu et al.).

**Current Major Geophysical Themes and Priorities:**

We highlight a number of geophysical topics that are central to larger outcomes:

**1. Subsurface oceans**, below ice shells, may represent the largest habitable niche in the galaxy by total volume. Such oceans may have minor external expression, or may remain largely hidden for decades to come. Despite this hidden nature, delegating study of them to future generations *may mean missing the lead in the overall story of extrasolar habitability*. Fortunately, our Solar System possesses numerous ocean worlds that inform this field: including Europa, Ganymede, Enceladus, Titan, Triton, Pluto, Charon, and large trans-Neptunian objects [1, 2]. NASA's recent programmatic Ocean Worlds focus may thus significantly benefit exoplanet research. Our Solar System does lack any ice-surface world larger than Ganymede, and also lacks known high-mass flux rate cryovolcanism, meaning that local studies must also be extended by modeling and by high-quality inference, to address worlds such as TRAPPIST-1 g & h, Proxima b, and many expected ice-surface planets to be confirmed in the future.

**2. Nomad worlds**, also known as rogue worlds, are a class of planets not gravitationally bound to host stars, and relate closely to the topic of subsurface oceans. Nomad worlds are primarily detected via microlensing surveys [3, 4] and are potentially habitable [5, 6, 7] by a range of mechanisms, most importantly water oceans below ice shells, supported by radionuclide decay. The statistics for terrestrial-size nomad worlds is unknown [8] but if early scattering of small worlds was efficient, could represent a significant fraction of total solid-surface bodies in the galaxy. It is important to recognize that beyond an ice line, habitability is controlled mainly by object size (and thus Uranium/Thorium content), and not by solar flux. Ice shell behavior, subsurface water ocean physics, and radiogenic modeling are focus areas for nomad habitability. *Support for microlensing surveys, including WFIRST, is critical for continued progress in this field*, as is support for high fidelity simulations of early solar system scattering rates.

**3. Plate tectonics** is widely regarded as having played an outsize role in shaping Earth's biological history, from its role in triggering entry into/escape from snowball Earth episodes, to flood basalt eruptions linked to mass extinctions, to the manner by which it controls the recycling of surface chemicals. Exoplanet research on plate tectonics [9, 10, 11] has revealed a complex range of possible outcomes that will only grow as further compositions are considered.

**4. Planetary densities**, derived from combined transit and radial velocity observations, can be a frustrating endpoint due to the degeneracy of compositions (largely between Fe/Ni, silicate, water, and H/He fractions) that can lead to common density results [12, 13]. In other cases, such as mini-Neptunes, composition can be better inferred [14, 15]. Steady progress in disk modeling, planetary formation modeling, and hydrodynamic simulations of impact erosion are key paths around this bottleneck. *Such work allows maximal scientific utilization of one of the main outputs of exoplanet observations: the mass-radius diagram.*

**5. Dynamical outcomes** depend on tidal feedback from interior states, through heavily studied but poorly constrained tidal quality factors [16]. Observations of eccentricity, spin, and obliquity, can thus contain information about internal melt fractions, asthenosphere presence, or the existence of any subsurface water ocean. *Interiors likewise critically influence terrestrial planet survival, in the face of eccentricity forcing and scattering [17, 18, 19], by controlling tidal damping.* Interiors play a key role in the process of tidal capture into spin-orbit resonances: It has been demonstrated for GJ 581 d [20] and GJ 667 Cc [21] that close-in exoplanets with a non-negligible eccentricity can, similarly to Mercury, *be tidally trapped in higher spin-orbit resonances.* This must influence climate, and can in principle shift a habitable zone. For a close-in exoplanet captured in a non-synchronous spin state, extreme tidal heat can change a planet's rheology and triaxiality, and can make the planet leave that rotation state and continue despinning down to either synchronism or, in some cases, pseudosynchronism (rotation rates a few percent faster than synchronous [22]), as may occur for the four innermost planets of TRAPPIST-1 (Makarov et al., submitted).

**6. Magnetic fields and high-pressure deep interiors,** directly influence observable phenomena such as atmosphere striping, mass loss, and mantle outgassing rates and composition. Understanding habitability, star-planet photochemistry, and mass-loss pathways to the bimodal radius distribution of exoplanets [23, 24], all depend centrally on magnetic fields, core crystallization, dynamo activity, and mantle convection rates. Exoplanet radio emissions can be used to measure magnetic fields [25] as cyclotron maser instability drives emission of radio waves at frequencies near the gyrofrequency [26]. The properties of a planetary magnetic field are often the least-constrained component of stellar-wind-induced atmospheric ion loss, such as for Proxima b [27], but are fundamental to understanding the space environment. Thus atmosphere detection for worlds undergoing severe loss [28], can lead to inferences about the deep-interior, such as composition, convective state, and energetic state. These observations provide a basis for inferring the condition of the dynamo region: a molten outer core (or other internal fluid layer), along with the likely buoyancy sources driving convection [29]. Such topics are poorly understood in the Earth-to-Neptune mass transition range where high pressure ices, superionic ice, and plasma ice phases, all interact. *Significant human capital, from at least the undergraduate level on up, will be strongly needed to address all of this geophysical complexity.*

**7. Exoplanet silicate volcanism** is highlighted in a separate whitepaper.

Note this brief list represents only a very small subset of exoplanet geoscience publications and subjects. Below, we place geophysics issues into the context of specific topics highlighted by the National Academy of Sciences 2018 Exoplanet Science Strategy solicitation:

**– Identify areas of significant scientific progress since publication of the New Worlds New Horizons Decadal Survey.**

- Detection of thermal variability on 55 Cnc e [30], possibly linked to geological activity.
- Recent spin dynamics research suggests 1:1 spin-orbit resonances may not be universal for close-in planets [20, 21, 22].
- The number of detected systems with complex, multi-body dynamics, such as TRAPPIST-1, Kepler-90, Tau Ceti, and 55 Cnc, (many at or near MMRs) has increased significantly.
- A clear bimodal distribution in exoplanet radii has been observed between the Earth and Neptune size range [23], with mass loss currently expected to play a role in its development.
- Evidence for water vapor plumes above Europa has emerged from HST observations [31], suggesting methods to chemically probe an exoplanet subsurface biosphere/cryosphere.
- The New Horizons spacecraft found young surfaces common in the Pluto-Charon system.
- Kepler catalog rocky bodies down to the mass of Mars, open an era of terrestrial world study.
- Proxima b, a terrestrial-class exoplanet, has been discovered at our nearest neighbor star, with unresolved questions of its surface being dominated by ice, ocean, or even lava [32].

**– Identify exoplanet science areas where significant progress will likely be made with current and upcoming observational facilities.**

- TESS and JWST combined will usher a revolution in understanding of small/terrestrial class worlds. Exoplanet science will move beyond only large planet characterization, to real validation of smaller terrestrial forward-modeling efforts.
- Improved resolution of the bimodal radius peak feature: does the radius valley depth reach zero? Are there similar features/bends at silicate/iron world transition points?
- The InSight lander will deliver seismology and core knowledge beyond the Earth-Moon system, *with an aspirational goal for seismology on a wide diversity of Solar System surfaces*.
- Possible first exomoons and exomoon population demographics, from TTV/TDV (While even sustained non-discovery would carry significant dynamical meaningfulness).
- TESS and new ground-based RV combined will lead to a significant increase in small-radius planets with measured masses and thus densities/bulk composition estimates. This is particularly relevant for planets around M dwarfs, with IR RV spectrographs coming online.
- Thermal phase curves may reveal the presence/absence of atmospheres on a handful of planets in the solar neighborhood, with possible detection of the 15 μm $CO_2$ band with this technique.
- Expanded constraints are needed to link atmospheres to surface types: such as by the prevalence of hazes, low/high molecular weight atmospheres in terrestrial worlds, aerosol differences between ice/silicate worlds, effects from changes in oxygen fugacity, and transitions between anoxic and oxic atmospheres.
- Doppler shifted reflected light and ionospheric disturbances in exoplanetary atmospheres may reveal the presence of large (magnitude > 9.0) seismic events, i.e., the range of tectonic activity *associated only with subduction*, a key fingerprint of plate tectonics [33].
- Density, eccentricity, and spin rate refinements (especially in multi-body systems), as well as concerted searches for exomoons and 55 Cnc e-analog planets, are top priorities for exoplanet geoscience.
- *The ability of WFIRST to conduct a 'statistical census of exoplanets with masses >$0.1M_E$ from the outer edge of the HZ to free floating planets' [34], by microlensing is essential to improve demographics for outer planetary systems. Retaining this mission is a very high priority for exoplanet geoscience.*

**– Identify exoplanet science areas and key questions that will likely remain after these current and planned missions are completed.**

- Modeling & indirect-inference will remain the key means to understand the *interiors* of world classes not represented in our Solar System, including worlds within the radius distribution bimodal notch, super-Earths, mini-Neptunes, Earth-size worlds with ice shells, deep water-ocean inundated worlds, magma ocean worlds, and exotic (e.g., C, Ca-rich) mantle compositions. Very large (GCM-level) modeling efforts will be required to address this tremendous compositional and parametric diversity.
- Galaxy-wide habitability may be dominated (by volume) by non-observable subsurface oceans. Modeling will inform if this is true, but conclusive answers will be elusive.
- The Ganymede-to-Earth-mass range of nomad worlds has a possibly very large population [35], with possibly frequent subsurface oceans, but full demographics may not be obtained with planned facilities. Detailed characterization will remain exceedingly difficult.
- Venus-analog worlds with dense clouds and haze will remain difficult to probe, given for example we are still not certain active volcanism is occurring on our neighbor.
- Only one of the four terrestrial planets in our inner Solar System would be well-characterized by transmission spectroscopy. Achieving direct imaging beyond nearby stellar systems presents a highly worthwhile challenge for hazy and solid-dominated worlds.
- Surface temperatures will be extremely hard to infer empirically from transmission and secondary eclipse data alone (assuming an atmosphere is present) even if the presence/absence of certain greenhouse gases can be confirmed.

**– Identify key observational, technological, theoretical, and computational challenges.**

- LUVOIR/HabEx/OST class observing capacity is critical for all exoplanet geoscience.
- *Filling the current exploration gap of the ice giants (Uranus and/or Neptune), is critical to the transition from super-Earths to mini-Neptunes and the bimodal exoplanet radius feature.*
- Europa Clipper & related ocean worlds programs (e.g., to Enceladus, Titan, Triton, or the trans-Neptunian region) are not only Solar System missions, but keys to overcome subsurface observing limits, to aim for a fuller accounting of galactic habitability.
- Modeling efforts must shift beyond a phase of 'assuming modern Earth-like starting conditions', *while at the same time study of the Earth's own complexity is far from complete.*
- Earth's Moon offers a cautionary tale that hydrated minerals can lead to an apparent spectroscopic signature for water, as observed by the EPOXI mission.
- Restrictions on theoretical end-member compositional cases are needed, such as what constitutes too much/too little $Fe/H_2O/P/S$/radionuclides/salts/anti-freezes, to allow or inhibit key phenomena? e.g., How much Fe is too much, if any amount, to inhibit plate tectonics?
- Solving how tectonics and outgassing work in super-Earths, with high fractions of primitive (reducing) materials will be challenging, and requires insight from outer Solar System bodies, including icy moons, comets, and water rich asteroids.
- Ongoing/future GCM efforts should be encouraged to include diverse highly volcanic cases.
- Global spin [36] could locate ring-of-fire type volcanic concentrations to infer plate tectonics.
- For key areas: plate tectonics, volcanism, tides, and volatile cycles, the diversity of possible chemistry is of enormous scope. This requires high-end computing, and large human capital needs. *Automation is necessary, but so is high individualized attention to key planets.*
- For exotic composition geochemistry and cycles, such as C & Ca-rich worlds, valuable small scale examples do exist on Earth to inform outcomes (e.g., carbonatite lavas in Africa [37]).

– **Discuss how to develop and expand partnerships (interagency, international and public/private) in furthering understanding of exoplanets and exoplanetary systems.**

• Advance access and use of exascale/mesoscale computing for planetary science. Example 1: migrating research in tidal dynamics into a more GCM-like paradigm, with lateral inhomogeneities handled via computing, not analytic models that require outdated simplifications. Example 2: mesoscale computing can determine ultra-high pressure material properties beyond the reach of diamond anvil efforts.

– **Identify likely fruitful cross-disciplinary topics and initiatives that will enable and accelerate progress in future areas of exoplanet inquiry.**

• Invite biology/biochemistry/biomedical fields into terrestrial world partnerships/settings, due to the central role of biochemical bottlenecks in creating Earth-analog outcomes.

• The geoscience community is moderately well integrated into exoplanet communities so far, but the need is rapidly growing, as *the field of 'exogeoscience' is set to expand at an explosive pace to match terrestrial planet discoveries by numerical-count.*

• However, to succeed, there needs to be greater research-level training in the earth sciences, both for astronomy students pursuing terrestrial exoplanetary science, and vice-versa.

• Planetary scientists are generally trained in analytic methods, not computer science. Partnerships with computer science can advance the use of mesoscale/exascale computing.

• Identify non-traditional elements/compounds essential to habitability and/or thermal evolution, to be quantified in protoplanetary disk observations (radiogenics, K, S, etc.).

• Use stellar element abundance observations (e.g., Hypatia catalog [38]) to constrain planetary structures. The stellar abundance community has insight into measuring hard (few and/or blend spectral lines) but especially relevant elements for internal heating like U, Th, & K.

• Whenever possible, breakdown/limit/prevent hard barriers and separated thinking between Solar System exploration and exoplanet science: using local examples will be the central means to advance exoplanet *interior* science for generations to come due to the unattainability of interstellar probes. The same is true for trans-Neptunian objects, as nomad world proxies.